\documentclass[aps,pre,onecolumn,superscriptaddress,showkeys,floatfix]{revtex4}
\usepackage{graphicx}
\usepackage{epsfig}

\newcommand{\Phot}{\textit{Photuris }}
\newcommand{\bs}{\symbol{92}}

\begin{document}

\title{Improved light extraction in the bioluminescent lantern of a \Phot firefly (Lampyridae)}


\author{Annick Bay}
\address{Research Center in Physics of Matter and Radiation (PMR),Department of Physics, University of Namur (FUNDP), 61 rue de Bruxelles, B-5000 Namur, Belgium}
\email{annick.bay@fundp.ac.be} 

\author{Peter Cloetens}
\address{X-ray Imaging Group, European Synchrotron Research Facility (ESRF), BP 220, 38043 Grenoble cedex, France}

\author{Heikki Suhonen}
\address{X-ray Imaging Group, European Synchrotron Research Facility (ESRF), BP 220, 38043 Grenoble cedex, France}

\author{Jean Pol Vigneron}
\address{Research Center in Physics of Matter and Radiation (PMR),Department of Physics, University of Namur (FUNDP), 61 rue de Bruxelles, B-5000 Namur, Belgium}

\begin{abstract}
A common problem of light sources emitting from an homogeneous
high-refractive index medium into air is the loss of photons by
total internal reflection. Bioluminescent organisms, as well as artificial
devices, have to face this problem. It is expected that life, with
its mechanisms for evolution, would have selected appropriate
optical structures to get around this problem, at least partially.
The morphology of the lantern of a specific firefly in the genus
\Phot has been examined. The optical properties of the different
parts of this lantern have been modelled, in order to determine
their positive or adverse effect with regard to the global light
extraction. We conclude that the most efficient pieces of the
lantern structure are the misfit of the external scales (which
produce abrupt roughness in air) and the lowering of the
refractive index at the level of the cluster of photocytes, where
the bioluminescent production takes place.
\end{abstract}


\maketitle
\section{Introduction}
Most living organisms - including ourselves - use light as a primary
source of energy and as an information carrier for communication.
Since the emergence of vision, light has kept delivering signals,
processed into knowledge and exploited at the behavioral level.
Vision and light flux management have turned into an essential
function for most highly evolved animals. Insects, birds, fishes,
and many other living organisms use spectrally filtered sun light
for making themselves conspicuous to their mates, warn predators of
toxicity, signal territory occupancy, hide from the view of preys or
predators and much more. 

At dusk, when the sun disappears from view, and when the direct sun
illumination vanishes, the importance of light does not weaken. Many
organisms have adapted to moonlight intensities and keep a clear
vision at night. Also, quite a few animals produce their own light
by bioluminescence and use it for communication in the dark. There
are two great categories of bioluminescent species: one in deep
waters, the other terrestrial. Bioluminescent animals from the deep
are usually emitting blue and get the help of symbiotic luminescent
bacteria \cite{Vezzi-sci-2005} to produce their light. In such an
association, a bacteria such as those in the genus
\textit{Photobacterium} takes care of the chemistry needed to
produce photons, in response to the host's needs. Unicellular algae
\textit{Pyrodynium bahamense} also produce light, especially in
disturbed waters. The \bs Phosphorescent Bay'' of Vieques, Porto
Rico, is a well-known place -- among many others -- where these
dinoflagellates can easily be observed producing a halo of blue
light following boat tracks.

\begin{figure}[t]
\centering\includegraphics[width=8 cm]{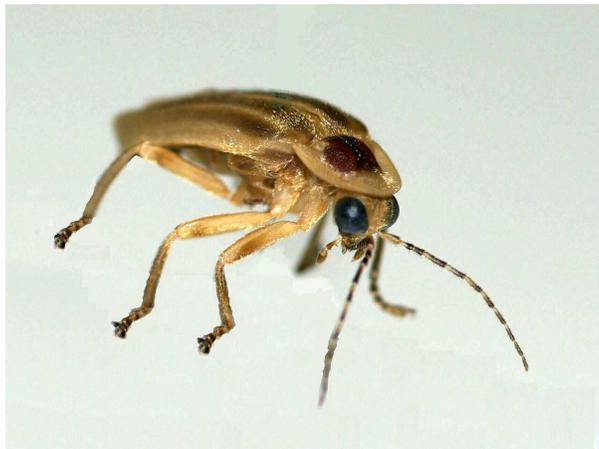} \caption{The
\textit{Photuris sp.} firefly, collected in the Darien forest,
Republic of Panama, whose lantern is investigated in the present
work.}\label{fig1}
\end{figure}

Some terrestrial animals such as insects also produce their own
light. The colors are most often in the yellow-green range, but
orange-red is also frequent. Perhaps the best known are fireflies
(see Fig. \ref{fig1}), in the family Lampyridae, for which all
phases, eggs, larvae, pupae and adults, males and females, can
produce light. Eggs and larvae produce a continuous light
emission. It is assumed that this illumination is a warning signal
against the predators, as in other steady bioluminescent insects
such as click-beetles (Elateridae: \textit{Pyrophorus}), which
start emitting a strong continuous signal when disturbed or
otherwise threaten. In some species, male and female adult
fireflies lanterns are more complex \cite{Ghiradella-mai-1998},
with the capability of flashing the bioluminescent emission. Each
species has its own specific light pulse pattern, which acts as a
communication signature: the female is usually at rest at the edge
of the forest while males are flying around, sending and receiving
signals. The females respond to \bs their'' males
\cite{Chapman-bk-1998} to reveal their location, and reproduction
is initiated. A perversion of this behavior is known, as the
females of the firefly \textit{Photuris versicolor} is able to
mimic the light flash pattern of some \textit{Photinus} females in
order to attract their males and prey on them
\cite{Lloyd-sc-1965}, absorbing nutrients and more
\cite{Eisner-pnas-1997}. The chemical reaction that creates
photons from the energy stored in adenosine triphosphate has been
described as the oxidation of a substrate of luciferine catalyzed
by the enzymatic luciferase. Different types of
luciferine-luciferase pairs have been described, leading to
different colors. What is not so well understood is the way the
bioluminescence can be switched on and off to produce flashes
that, sometimes, can be short (pulse duration under a second) and
undergo amazingly fast rises and falls.

Another problem -- which has has been neglected up to now, except
maybe for a preliminary report of part of the present work in a
conference abstract \cite{Bay-SPIE-2009} -- is the optical origin of
the \textit{external} efficiency of the firefly lanterns. When a
source emits from inside a high refractive index medium, the emitted
light that reaches the surface will not always be able to escape
into the air, because of the total reflection at large incidence
angles. The external efficiency of all electroluminescent devices
made from semiconductors \cite {Benisty-bk-1999} is affected by
this effect, and this is especially true for the recently developed
large-size organic light-emitting diodes
\cite{Vandersteegen-6486-2007,Vandersteegen-6655-2007}. The firefly
lantern is no exception and it is interesting to parallel the
decades of semi-conductor diodes structural optimized engineering
with some hundred million years of blind natural selection in these
living organisms.

The present work considers the photonic structure of the various
parts of the insect's lantern in order to recognize, on the ground
of computer simulations, those who plausibly contribute best to the
improvement of the light extraction.

We will carry out this work in five steps : (1) [section \ref{flat}]
a reference system is first defined; (2) the results of morphology
investigations is reported; (3) computer simulations of the optical
behavior of each individual recognized substructural lantern element and
conclusions on the positive or adverse effect on light extraction
are presented; (4) a global model, that takes into account the set of relevant substructures, will be considered in order to
provide some theoretical estimate of the external efficiency; (5)
finally, we show how these calculated estimates get support from
direct observations.

\section{Light extraction from an homogeneous medium terminated by a
planar surface}\label{flat}

We do not know how fireflies looked like some hundred million years
ago, so it is difficult to use the ancestors' extraction organs as a
reference to which compare the lanterns that are around today, in
nature. Our reference, in order to discuss possible improvements,
will be based on an idealized physical model system consisting of an
homogeneous medium in contact with air at a planar surface. We will
set an isotropic point light source at some macroscopic distance
under a perfect flat surface and evaluate the fraction of the
emitted light intensity that effectively crosses the surface to be
released in the open air (in the emergence medium hemisphere). This
fraction will represent the external efficiency reduction factor.
All subsequent calculations, when the structured parts of the
firefly lantern will be included, will be treated in exactly the
same way.

\begin{figure}[b]
\centerline{
\includegraphics[width=10.0 cm]{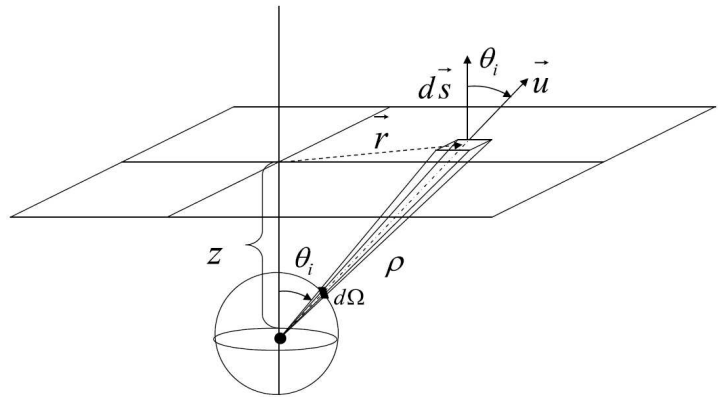}}
\caption{Geometric data for the integration of the transmitted light
over the surface of the reference homogeneous system.} \label{fig2}
\end{figure}

Shown in Fig. \ref{fig2}, a point source emits uniformly at a
depth $z$ in a medium of refractive index $n_i$, limited by a
plane surface. It illuminates this surface in a non-uniform way.
If a sphere with unit radius is drawn, centered on the source, the
total power emitted through its upper hemisphere is $P_0/2$, where
$P_0$ is the total radiant energy emitted each second by the
source. Assuming that there is no absorption between depth z and the surface, the fraction of this power sent out in a solid angle
$d\Omega$, under the incidence angle $\theta_i$ is then
\begin{equation}
\frac{{d\Omega }}{{2\pi }}(P_0/2)  = \frac{{d{\kern 1pt} \vec s
\cdot \vec u}}{{2\pi \rho ^2 }}(P_0/2)  = (P_0/2) \frac{ds\;{\cos
\theta _i }}{{2\pi \rho ^2 }}
\end{equation}
where: $d \vec s$ is the element of horizontal plane surface
intercepted by the solid angle $d\Omega$; $\vec u$ is a unit
vector in the light propagation direction with incidence angle
$\theta_i$;  $\rho$ is the distance between the source and the
surface element.

The horizontal surface element can be expressed in polar
coordinates $r$ and $\phi$, $ds = r\,dr\,d\phi$ to give (with $r =
z \tan{\theta_i}$)
\begin{equation} \frac{{d\Omega }}{{2\pi
}}(P_0/2)  = \frac{{(P_0/2) }}{{2\pi }}\sin \theta _i \,d\theta
_i\,d\phi
\end{equation}
The total power incident on the plane surface can be obtained by
integrating this contribution over all surface elements, which
leads us to the expression
\begin{equation}
P_i = (P_0/2) \int_0^{\frac{\pi }{2}} {\sin \theta _i \,d\theta _i
}
\end{equation}
This, as expected, turns out to be $P_0/2$. The power transmitted
into the outer medium, with a refractive index $n_t$ depends on
the transmission coefficient $T \left(\theta_i\right)$ of the
interface for each incidence $\theta_i$. The total power
transmitted can then be obtained as
\begin{equation}
P_t = (P_0/2) \int_0^{\frac{\pi }{2}} T \left(\theta_i\right)
{\sin \theta _i \,d\theta _i }
\end{equation}

The fraction of the incident energy transmitted in the outer
medium is then given by
\begin{equation}\label{fraction}
\frac{P_t}{P_i} = \int_0^{\frac{\pi }{2}} T \left(\theta_i\right)
{\sin \theta _i \,d\theta _i }
\end{equation}
This result is independent of the depth $z$ from which the light
is emitted, due to the lack of a naturally defined length scale
for an homogeneous material under a flat surface.

The transmission coefficient for a planar interface that separates
two dielectric media is given explicitly by Fresnel's formula's.
For unpolarized light, we have the average of transverse electric
and transverse magnetic transmissions :
\begin{equation}
T\left( {\theta _i } \right) = \frac{1}{2}\left\{ {\frac{{\sin ^2
\left( {\theta _i  - \theta _t } \right)}}{{\sin ^2 \left( {\theta
_i  + \theta _t } \right)}} + \frac{{\tan ^2 \left( {\theta _i  -
\theta _t } \right)}}{{\tan ^2 \left( {\theta _i  + \theta _t }
\right)}}} \right\}
\end{equation}
where the angle of emergence $\theta_t$ is obtained from Snell's
law of refraction,
\begin{equation}
\sin \theta _t  = \frac{{n_i }}{{n_t }}\sin \theta _i
\end{equation}

These expressions can be used in the integral (\ref{fraction}) in
order to obtain the total transmission, integrated over all
incidence angles $\theta_i$. In the reference system we consider,
we require $n_i$~= 1.56, which is the refractive index found in
hard parts of the insect's cuticle \cite{Sollas-rsl-1907,Vukusic-rsl-99,Yoshioka.2011}, and $n_t$~= 1, the refractive
index in air. Since the refractive index of the incidence medium
is larger than the refractive index of the emergence region, we
will have total reflection for incidence angles $\theta_i$ larger
than the critical angle $\theta_c$ given by
\begin{equation}\label{critical}
\sin \theta _c = \frac{{n_t }}{{n_i }}
\end{equation}
The transmission coefficient $T \left(\theta_i\right)$ vanishes
for $\theta_i > \theta_c$ and the integration (\ref{fraction})
needs not be carried out further above this limit.

The integral in (\ref{fraction}) can easily be carried out
numerically, to any needed accuracy, using a variable-step routine
\cite{Forsythe-bk-1977}. For the homogeneous medium of refractive
index of 1.56, we find a global transmitted fraction of 20\%.

This result may seem surprisingly low, especially for a medium
with a very moderate refractive index, such as the one considered
here. It should be noted that with these refractive indexes, the
critical angle for total reflection is 40$^{\circ}$, as given by
(\ref{critical}), so that none of the incidence angles $\theta_i$
between 40$^{\circ}$ and 90$^{\circ}$ contribute to the
transmission. At small angles, for $\theta_i$ near 0$^{\circ}$,
the geometric factor $\sin{\theta_i}$ also cuts the transmission
efficiency, even if the Fresnel's transmission coefficient is
strong there (about 95\%).

This 20\% transmission will be our reference in the models we will
be developing in the next sections. But before coming there, we
will describe the structures in which the emitted light has to
propagate, from the air down to the photocytes (the light-emitting
cells), outside the cuticle surface.

\section{Morphology of optical media in the firefly lantern}
\label{morphology}

Morphological studies have been carried out on a \textit{Photuris} firefly. Details on collection of the specimens, the determination of the genus, as well as the sample preparation, used analyzing techniques and simulations tools are given in the section "Materials and Methods" at the end of this paper.

\begin{figure}[b]
\centerline{
\includegraphics[width=11.3 cm]{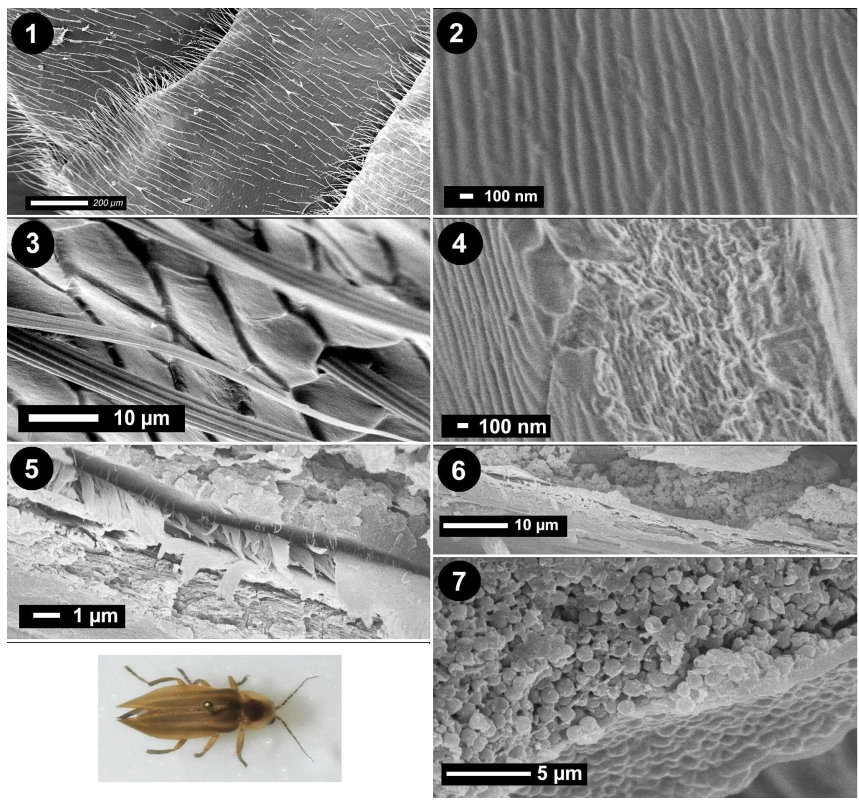}}
\caption{Observed structures crossed by the emitted light in the
lantern of the firefly. (1) Setae; (2) ribs separated by 250 nm;
(3) misfit scales, 10 $\mu$m long and 3 $\mu$m high; (4) textured
cuticle volume, 2.4 $\mu$m thick. The insert is a top view of the
structure; (5) muscular layer (tentative interpretation), variable
thickness; (6) photocytes cluster membrane, 60 nm thick; (7)
photocytes layer, containing a high density of spherical
peroxisomes.} \label{fig3}
\end{figure}

\begin{figure}[t]
\centerline{
\includegraphics[width=12.0 cm]{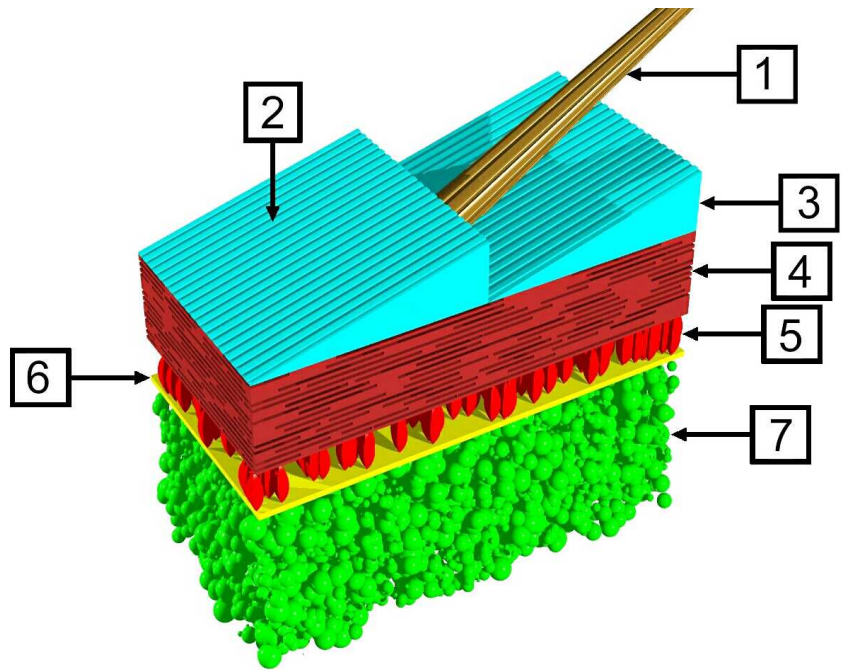}}
\caption{The various structures to be examined for light
extraction improvement. (1) Setae; (2) ribs separated by 250 nm;
(3) misfit scales, 10 $\mu$m long and 3 $\mu$m high; (4) textured
cuticle volume, 2.4 $\mu$m thick; (5) muscular layer (tentative
interpretation), variable thickness; (6) photocytes cluster
membrane, 60 nm thick; (7) Photocytes layer, containing spherical
peroxisomes.} \label{fig4}
\end{figure}
Typical SEM images of these structures are shown in Fig.
\ref{fig3}. In order to facilitate the description of the observed
features, Fig. \ref{fig4} summarizes the observations gathered on
the structure of the luminescent abdominal segments, on the
ventral side of the insect. Seven photonic sub-structures have
been identified, which have a potential impact on the light
extraction efficiency.

\textbf{Substructure (1).} The setae found on the cuticle are
rigidified by longitudinal ridges, as often in arthropods. These
straight, conical projections are likely part of sensitive organs.
They can conduct light and scatter it out, but they are not
numerous and their cross-section (5 $\mu$m in diameter) is small
compared to their nearest neighbors' distance (50 $\mu$m).

\textbf{Substructure (2).} The cuticle surface is divided into
scales -- see structure (3), below -- and the scales surface show
roughness arising from a fine grating of parallel ribs. The ribs
are separated by a distance of 250~nm and are typically protruding
100~nm above the scale surface. The cross-section profile of these
ribs is very smooth.

\textbf{Substructure (3).} The cuticle of the abdominal segments
is composed of scales joined together at their perimeter. However,
contrasting many other Coleoptera, the scales do not constitute a
flat pavement, but one of the edges (towards the abdominal tip) is
slightly protruding. This misfit between scales gives rise to an
abrupt running edge which can contribute to light scattering and,
through this mechanism, impacts the overall light extraction. The
average distance between the protruding edges rows is about
10~$\mu$m, long compared to the emitted light wavelength, but
surprisingly small, compared to the usual size of Coleoptera
scales. The height of the protrusion is variable, but 3 $\mu$m is
a frequent value.

\textbf{Substructure (4).} Just below the scales, we note a layer,
2.4 $\mu$m thick, which makes the bulk of the tegument in the
abdominal segment. This layer is a stack of about 30 plates, each
80 nm thick. Each plate is composed of an homogeneous chitin slab,
30 nm thick, bearing a network of parallel rods which act as
spacers for the next plate. The spacers corrugation forms a
regularly spaced grating with a profile close to that found in the
ribs, on the scale surface. The distance between the spacers is of
the same order as that of the surface grating, 250~nm. The height
of the corrugation associated with these spacers is 50~nm.

\textbf{Substructure (5).} The next inner layer is seen to be
essentially empty, except for a large number of fibers which bind
to the neighboring layers, (4) and (6). It is difficult to provide
data on the thickness of this layer because it is found to be
variable over a very wide range. Many different samples provide
very different values and it is plausible that this gap between
solid layers is tunable by muscular contractions.

The fibers found there may then be some remains of muscles, that
can control the geometry of the gap. This interpretation is still
very tentative but, if correct, it may be important for the
understanding of the production of very fast pulses by the
firefly. The very fast \bs on'' and \bs off'' switching of the
bioluminescent chemical reaction that produce light pulses is
still unexplained because of apparent timing inconsistencies. In
particular, the fast admission of oxygen to start the reaction and
the fast injection of an inhibitor to stop it cannot be accounted
for by fluid diffusion, considered too slow. A strong pump
mechanism would be needed, and this may be provided by the action
of muscles that act on the reaction chamber directly, taking its
bearing on the mechanically stiff layer (4). Insect muscles can be
assumed to be fast and strong enough to produce the fast fluid
motion needed to control the emission switch. Further work is in
progress to assess the real value of this suggestion, which
should, at the moment, be taken with care.

\textbf{Substructure (6).} The gap described in (5) opens on a
solid, flexible membrane, 60 nm thick, which separates this \bs
fibrous'' gap from the chamber (7) where the bioluminescent
reaction takes place.

\textbf{Substructure (7).} The reaction chamber is found below the
above layers. It appears to be filled with granular bodies, close
to spheres: the peroxisomes. Peroxisomes have been recognized as
organelles by the cytologist Christian de Duve in 1969
\cite{Duve-sc-1969}. One of the main function of these organelles
is to eliminate toxic peroxides. They contain a crystalline core
and are known to incorporate high quantities of urate oxidase and
other enzymes. Figure \ref{fig4} shows a particularly large quantity
of peroxisomes in the layer where the bioluminescence is taking
place. These granules, with an average diameter of 0.9 $\mu$m,
appear on this SEM image to be densely packed, providing a
macroporous disordered structure.

\begin{figure}[b]
\centerline{
\includegraphics[height=4.2cm]{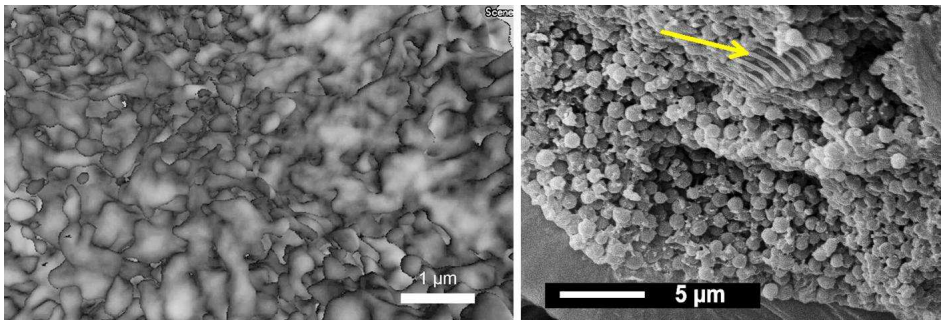}}
\caption{(a) X-ray nanotomography landscape image, extracted from
a three-dimensional representation of the internal structure of a
photocyte. These three-dimensional data confirm a disordered and
slightly multi-dispersed distribution of the peroxisomes,
suggesting an isotropic diffuse light emission. (b) Spherical
peroxisomes in the photocytes. Note the remains of a fractured
tracheole (arrow) which brings oxygen for the luciferine
oxidation.} \label{fig5}
\end{figure}

This disordered structure has been confirmed by an X-ray
nanotomography experiment conducted at the European Synchrotron
Radiation Facility (ESRF) in Grenoble (France) (see \bs Materials and Methods'' section  at the end of this paper). This technique does not require the lantern fracturing to observe the volume disorder of the lantern. This presents a huge advantage in comparison to SEM which is a more invasive technique. Three-dimensional
data describing the firefly photocyte's was generated, with about
50 nm pixel size. It is found that the peroxisomes assume a
continuous range of diameters, from 0.3 to 1.5~$\mu$m. The disorder is not only a disorder of sphere positions but also a disorder in the size of the spheres. This increased disorder reinforces the argument that the photocyte medium optically behaves as an homogeneous material. The lower refractive index of this averaged medium impacts light extraction by slightly changing the critical angle for total reflection. No directional correlation is found in the spherule arrangement,
implying that the peroxisomes distribution avoids both long-range
and short-range orders (see Fig. \ref{fig5}(a)). This information
suggests that the peroxisomes arrangement cannot drive the emitted
light preferentially normal to the cuticle surface as ordered
photonic crystals can plausibly do \cite{Vukusic-sc-2005}. The
multiple scattering of light in this strongly disordered structure
is primarily expected to evenly diffuse the emitted light in all
directions. This provides support for modeling this isotropic
region as a homogeneous medium with an averaged refractive index.

The filling factor of this structure can be estimated to be close
to 75\%, similar to that of periodic hard-sphere compact
structures. Figure \ref{fig5}(b) shows the assembly of spherical
granules together with the remains of a fractured tracheole,
confirming that the structure is indeed the place where the
oxidation reaction takes place: the tracheoles drive oxygen to the
porous structure when sustaining the bioluminescence reaction that
produces light.

\section{Calculation of light propagation in the lantern substructures}
\label{computer}

\begin{figure}[b]
\centerline{
\includegraphics[height=4.5cm]{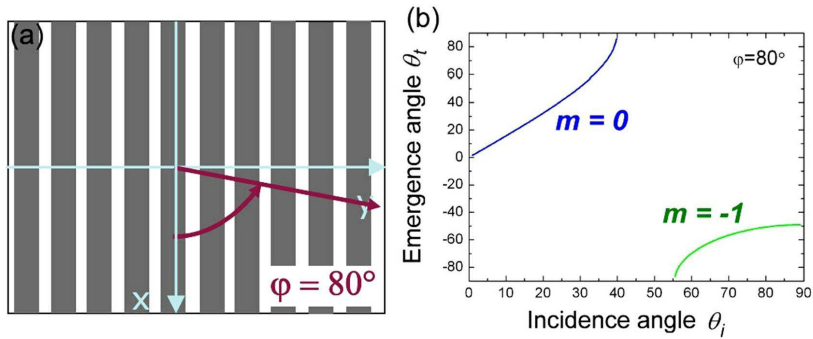}}
\caption{Diffraction by the 250 nm grating formed by the
(a) longitudinal ribs on the surface of the firefly abdominal scales for a fixed incident azimuthal angle $\phi=80^\circ$. The diagram in (b) results from application of Eq. (\ref{equ-9}). This shows the existence of diffracted orders (m=-1), but does not provide any information about their intensities. These are calculated and reported in Fig. \ref{fig7}.}
\label{fig6}
\end{figure}

We will now reconsider all the substructures revealed
by the scanning electron microscopy analysis as described in the previous section and, on the basis of
computer simulations, investigate the optical properties of each observed substructure separately in
order to discuss their possible contribution to light extraction.

For propagation calculations, we solve Maxwell's equations, which
means that we use a fully vectorial representation of the light
waves and we account for multiple scattering at all stages of the
calculations. We use transfer-matrix techniques
\cite{Vigneron:2006} to calculate transmissions through both
one-dimensional and two-dimensional photonic-crystal films (see \bs Materials and Methods'' section). 

The light emitted by the firefly is a yellow-green color of narrow
spectral range. A wavelength of 560 nm is adequate to represent
the color of the flashes and all the simulations in this section
will be carried out for this specific wavelength.

The angular distribution of the incident light is kept identical
to that defined in the reference model. The different light rays
impinging on the surface under different directions are considered
incoherent. It should be noted that these rays represent radiation
arising from an extended source (the cluster of photocytes) whose
different parts emit incoherently.

In the following we use for all relevant substructures diffused intensity diagrams that describes the integrated intensity of the light scattered in the whole emergence hemisphere as a function of the incident polar and incident azimuthal angle ($\theta$ and $\phi$). These diagrams are important to appreciate the contribution of each substructure to the light scattered out of the medium, especially for incidences in the total reflection conditions of the reference system.

\textbf{Substructure (1).} The cross section of the setae
represent less than 1\% of the surface offered to light transit so
that, even if they are expected to deliver much of the light they
carry, their contribution to the external efficiency will be low.
They are probably not dense enough to have a strong positive
effect on the light extraction. However, they should not show any
adverse effect.

\textbf{Substructure (2).} The ribs on the scales form a grating
with a period $b = 250$~nm (see Fig. \ref{fig6}(a)). This lattice
does produce diffraction, which allows to use some of the light
which, otherwise, would have been lost in the total reflection.
The structure produces both refraction (transferring light from a
medium of refractive index $n_i$ into a medium with refractive
index $n_t$) and diffraction, changing the tangential wave vector
component by $m (2 \pi / b)$ in the direction normal to the ribs.
For light impinging at right angle with the ribs, for instance,
the emergence angles $\theta_t$ are related to the incidence angle
$\theta_i$ through the Eq. (\ref{equ-9})

\begin{equation}\label{equ-9}
\sin \theta_t  = \frac{{n_i }}{{n_t }}\sin \theta_i  +
m\frac{{\left( {{\lambda  \mathord{\left/
 {\vphantom {\lambda  {n_t }}} \right.
 \kern-\nulldelimiterspace} {n_t }}} \right)}}{b}
\end{equation}
For instance (see Fig. \ref{fig6}(b)), for an angle of incidence
$\theta_i = 60^{\circ}$, well within the total reflection region,
the order $m = -1$ produces a transmitted beam which escapes under
$\theta_t = -62.7^{\circ}$ (backscattered). When the azimuthal
angle $\phi$ is changed, the angle of emergence is increased
(grazing backscattering) and the diffraction disappears for
incidence planes close to parallel to the ribs.

\begin{figure}[b]
\centerline{
\includegraphics[width=7.5 cm]{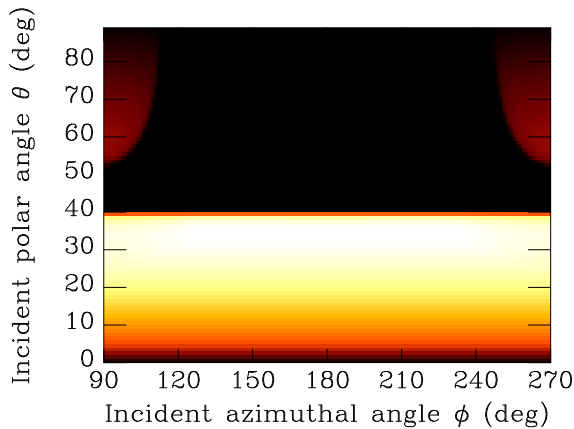}}
\caption{Hemispheric transmission (logarithmic scale) of a
sine-square grating with step $b=250$~nm, as a function of the
incidence angles (polar and azimuthal). The geometric factor
$\sin{\theta_i}$ is included. See Eq. (\ref{fraction}). The
intensities above $\theta_i = 40^{\circ}$ are diffracted.}
\label{fig7}
\end{figure}

This, however does not inform us on the impact of diffraction on
the efficiency of the light extraction. We must calculate the
intensity of the transmitted beams ($m = 0$ and $m=-1$) and
integrate these on all incidence angles before reaching any
conclusion. This has been done with a model where the profile is
approximated by a simple sine square function
\begin{equation}\label{equ-10}
z\left(y\right)=h \, \sin^{2}{\left(\frac{\pi}{b}\,y\right)}
\end{equation}

Eqs. (\ref{equ-9}) and (\ref{equ-10}) only address relationships between the incident and emergent polar angles and are completely independent of the polarization state of the incident light. They do not provide information about scattering intensities which actually depend on the polarization.

The distribution of hemispheric intensities as a function of the
incident (polar and azimuthal) angles $\theta_i$ and $\varphi_i$
are shown in Fig. \ref{fig7}. Incident azimuthal angles are counted in the x-y plane (see Fig. \ref{fig6}). For symmetry reasons, we account only for the half hemisphere. The polar angle $\theta$ is defined from the direction perpendicular to the interface, whereas the azimuthal angle $\phi$ is defined in the plane of this interface. Azimuthal angles of 90$^\circ$ and 270$^\circ$ correspond to incidences perpendicular to the ripples, whereas an azimuthal angle of 180$^\circ$ corresponds to an incidence parallel to these ripples.

We note that, except for the low diffracted intensities for polar incident angles larger than 40$^\circ$ the hemispheric transmission is very close to that expected from a planar surface. 
The integration of this distribution
gives a total transmission of 20.65\% for the TE incident
polarization and similar numbers for TM incident polarization (see "Material and Method" section): the corrugation does not improve significantly on
the initial 20\% transmission of the flat surface. This
corrugation is too weak and the grating period is too small to
bring any significant change in the light extraction.

\textbf{Substructure (3).} The misfit of scales gives a larger
size corrugation, and this gives more opportunities to produce
diffracted beams of large order. In calculating the effect of this
corrugation, we must ensure that the calculation has converged as
a function of the number of diffraction orders included.

\begin{figure}[b]
\centerline{
\includegraphics[width=7.5 cm]{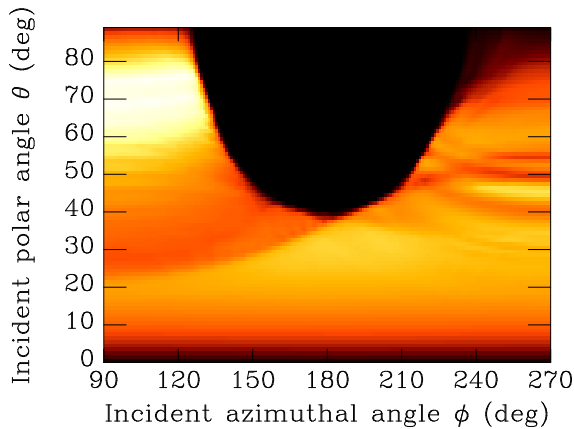}}
\caption{Hemispheric transmission (logarithmic scale) of a grating
with a triangular profile, with step $b=10~\mu$m, as a function of
the incidence angles (polar and azimuthal). The geometric factor
$\sin{\theta_i}$ is included. See Eq. (\ref{fraction}).}
\label{fig8}
\end{figure}

The model consists of an array of oblique tiles 10~$\mu$m long,
slanted in such a way that one of their side is protruding
3~$\mu$m out of the surface. This gives a sharp edge, repeated
every 10~$\mu$m, giving a large-step grating. The distribution of
the transmitted light, including all diffracted orders, is shown
in Fig. \ref{fig8} as a function of the polar and azimuthal
incidence angles. An azimuthal angle of $\phi=90^\circ$ corresponds to an incidence perpendicular to the sharp
edges, $\phi=180^\circ$ corresponds to an incidence parallel to the protruding edges and $\phi=270^\circ$
describes an incidence perpendicular to the slope of the edges

The result is quite different from that produced
by a short-period grating as the one considered in the
substructure (2). The directions, above $\theta_i = 40^{\circ}$
for which total reflection was expected are now strongly
transmitting. The optical phenomenon is actually better described
as light scattering on the edges of the tiles, as separated
diffraction beams are not perceived.

The origin of this redistribution of light is then qualitatively
different from the the mechanism mentioned in the ribs grating (2)
above, because of the edges separation (10~$\mu$m). The integrated
transmission also reaches a more interesting value : 29.5\% (an
average between the results for TE incident polarization, 29.3\%
and TM, 29.6\%, with 32 diffraction orders). The scattering by the
scale edges, with the large maintained misfit, improves very
significantly the light extraction, adding 47.5\% more light out.

\begin{figure}[t]
\centerline{
\includegraphics[width=7.0 cm]{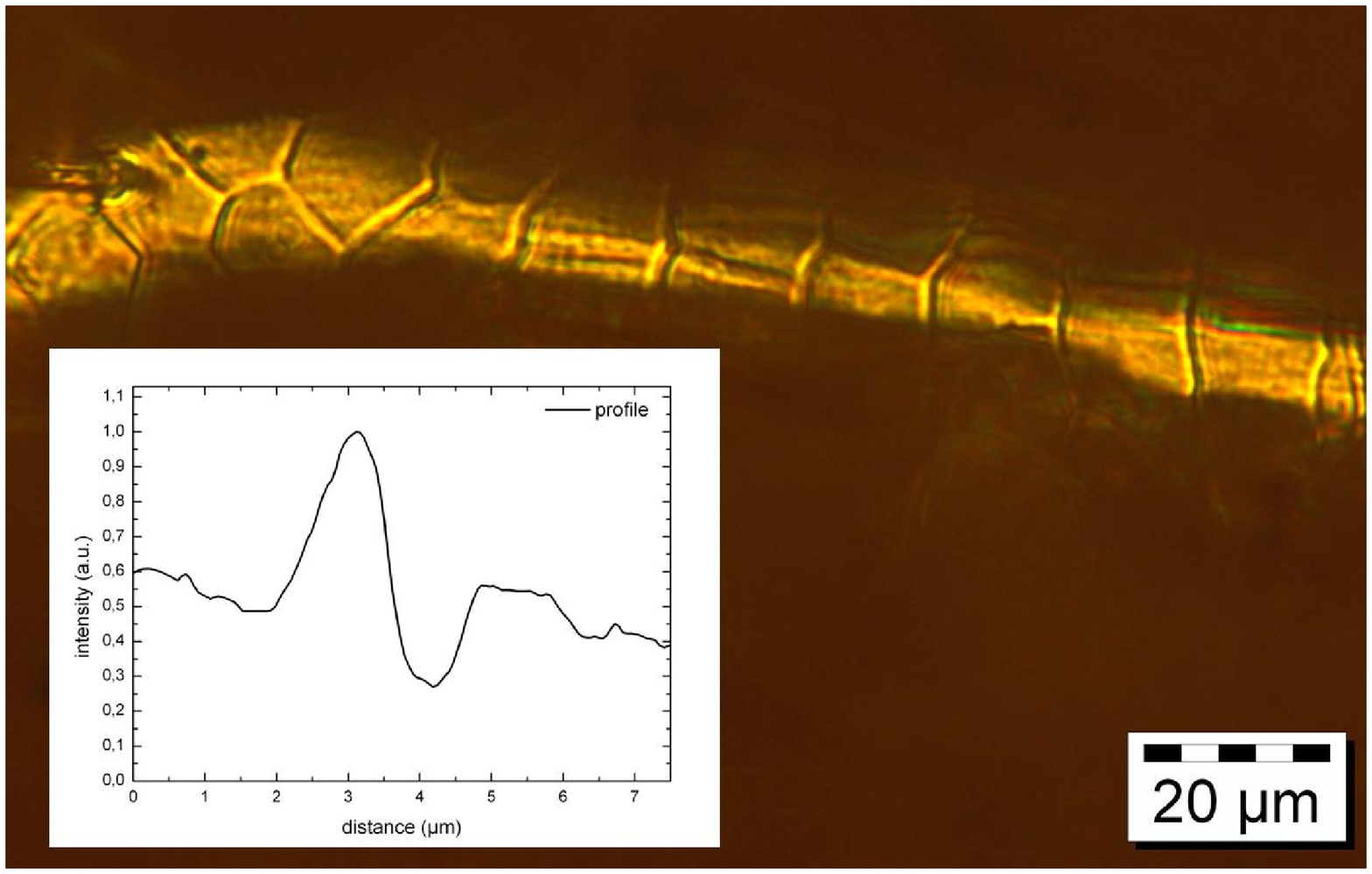}}
\caption{Light transmission image of the scales covering the
surface of one of the firefly's light-emitting segments, viewed in
an optical microscope with illumination set for transmission
through the sample. The high-intensity trace concentrated on the
edges of the protruding scale adds to the transmitted intensity
and favors light extraction. The incrustation shows an intensity
profile across the scale edge, in the image plane.} \label{fig9}
\end{figure}

The significant increase of light transmission at the protruding
edges of the scales is confirmed by the observation of the
non-uniform transparency of the cuticle. Figure \ref{fig9} clearly
shows the higher light intensity exiting from the scale's edge,
when the cuticle is illuminated from the photocyte location and
observed from the outside of the abdomen. In order to obtain this
image, an Olympus BX61 optical microscope (objective 60X) was used
in transmission mode. The incrustation in Fig. \ref{fig9} shows an
intensity profile across the scale edge, in the image plane. The
maximum intensity is obtained on the protruding side of the edge,
while a darker fringe develops on the lower side.

\textbf{Substructure (4).} This structure is the repetition, 30
times, of a double layer 80~nm thick, made of 30~nm of solid
chitin, and a 50~nm inhomogeneous layer that can be treated with
the sine-square model described by Eq. (\ref{equ-10}). The
lateral period of this sine square is $b = 250~$nm, as for the
surface ribs grating. We have seen already that this period
provides very few diffracted waves (order $m = -1$). The
calculation of the integrated hemispheric transmission can be
carried out as before, giving 20.7\% as a result. This value, as
for the surface ribs corrugation, is not significant and we
conclude that the volume corrugation of the cuticle has little
(but also, no adverse) effect on the light extraction. The
function of this inhomogeneity may be mechanical stiffness.

\textbf{Substructure (5).} The \bs muscular'' layer is difficult
to optically characterize, as the contents and thickness range are
not precisely known. With a thickness as small as 90~nm, frequent
on SEM observation, its effect on the light transmission is weak.
It will simply be included later in a more global model before
bringing our conclusions from this study.

\textbf{Substructure (6).} This membrane is extremely thin (60~nm
on the average) so that its effect on the light transmission is
very limited. It will also be included in a more global model
below, but we will not consider it separately.

\textbf{Substructure (7).} The optical structure of the reaction
chamber, with the accumulation of peroxisomes, is another
interesting structure, with no equivalent, to our present
knowledge, in other insects. The confinement of the chemical
reactants certainly has some influence on the internal quantum
yield \cite{PhysRevB.64.201104}, which depends on the photon
density of state, which is highly dependent on the inhomogeneous
refractive index distribution. The internal efficiency of the
bioluminescent reaction turns out to be close to 88\%
\cite{Vico-jpp-2008}. The average refractive index found in the
chamber is, however lower than that of chitin, because of the
presence of the spherical peroxisome : urate, with a refractive
index of 1.4, produces a significant lowering of the \bs incident
medium'' refractive index (containing the source) and this
lowering must lead to a displacement of the critical angle for
total reflection. Assuming a regular hexagonal compact assembly of
the peroxisome spheres, with a filling factor of 0.74, one can
estimate an average refractive index of 1.38, assuming a lower
(1.33) refractive index in the small interstitial volume. The
critical angle, as given by Eq. (\ref{critical}) turns out to be
46.4$^{\circ}$ in that case. This supplement of 6$^{\circ}$
compared to our homogeneous reference opens a new track for light
escape and improves the overall external yield.

\section{Global model}\label{global}

\begin{figure}[b]
\centerline{
\includegraphics[width=7.5 cm]{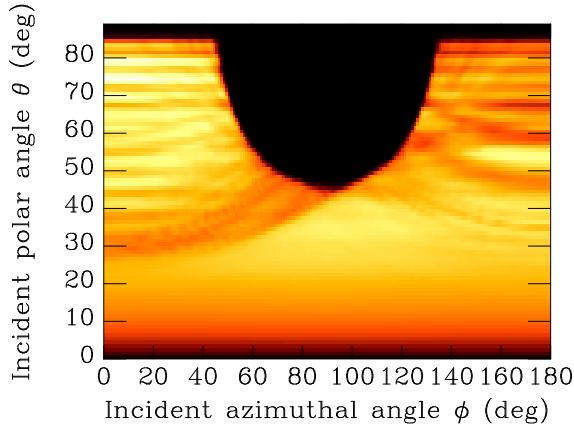}}
\caption{Hemispheric transmission (logarithmic scale) of the whole
structure, as a function of the incidence angles (polar and
azimuthal). The geometric factor $\sin{\theta_i}$ is included. See
Eq. (\ref{fraction}).} \label{fig10}
\end{figure}

We have estimated the gain by considering a more global model, in
which all constituents propagating the light are included
together: the incident medium is homogeneous, with a refractive
index $n_i = 1.38$, separated from a \bs muscular'' gap of 90~nm
by a thin membrane (60~nm) with a refractive index 1.56, a
2.4~$\mu$m thick cuticle layer with an average refractive index of
1.38 and a surface terminated by a triangular profile representing
the 10 $\mu$m scales. The average refractive index of 1.38 was
calculated by a procedure similar to that used for averaging the
dielectric constant for the coloring structure of the blue beetle
\textit{Hoplia coerulea} \cite{Vigneron-pre-hop-05}, indeed very
similar to the structure found here, except for the layer
thicknesses. The ribs and the setae on the surface of the scales,
at the outer surface, are neglected because they were shown,
above, to only contribute weakly to the light extraction.

The photocytes cluster (incident region) is considered
homogeneous, which means that we neglect diffraction by this
medium. This is justified by the moderate difference of refractive
index  between the peroxisome spheres and the fluid interstitial
medium, but also because of the isotropic spherules organization.

The transmission for all incidence angles is shown in Fig.
\ref{fig10}. Due to multiple scattering, we obtain a rather
uniform angular distribution of the light, except for
$\varphi_i~=~0$ or $\varphi_i~=~180^{\circ}$, if $\theta_i$
exceeds the critical angle of 46$^{\circ}$. These azimuthal angles
correspond to light impinging on the surface grating in the \bs
long'' direction, where the grating experiences a total
translational invariance. In that direction, we recover the
flat-surface condition and the concept of a critical angle
reappears. We should note that the strictly two-dimensional
grating, with one invariant direction, is an approximation. The
scales edges are not exactly rectilinear and the scales area is
actually limited in two directions.

The integration of the transmission into the full emergence hemispheric solid angle for all incidence
angles leads, for this global model, to a transmission of 39.6\%,
to be compared to the 20\% found for the homogeneous medium with a
planar surface. This value is for unpolarized light, the average
between 37.9\% for TE waves and 41.3\% for TM waves. The fraction
of the enhancement specifically due to the lowering of the
incident medium refractive index is of the order of 5\%.

\section{Support from direct optical measurement}

\begin{figure}[b]
\centerline{
\includegraphics[width=7.5 cm]{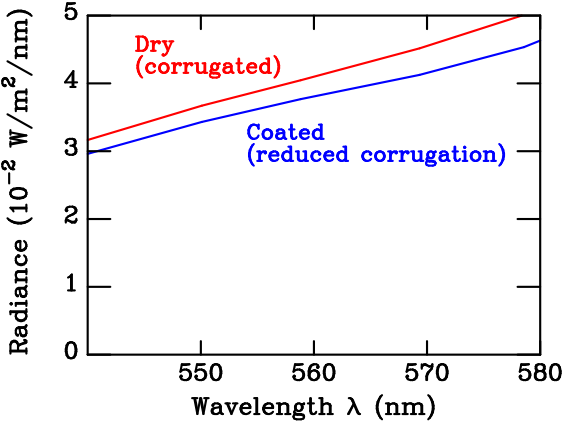}}
\caption{Hemispheric radiance of the abdomen of a
\textit{Photinus} firefly, with (red) and without (blue) its
surface corrugation. The suppression of the corrugation, in the
latter case, was obtained by coating the ventral part of the
abdomen segments with a refractive-index matching liquid.}
\label{fig11}
\end{figure}

An experimental check of these theoretical results could be
carried out by measuring the hemispheric radiance of the firefly
ventral side of the abdomen, while internally illuminating the
photocytes. The abdomen of a Photuris firefly was separated from
the thorax and an optic fiber, connected to a white halogen light
source, was introduced along the antero-posterior direction,
through the thorax/abdomen section. The radiance (W/sr/m$^2$/nm)
of the ventral side was measured in a solid angle spanning all
azimuthal angles and polar angles from 0 to 80$^\circ$ in a ELDIM EZ Contrast XL80MS
scatterometer (see \bs Materials and Methods''), for wavelengths close to the bioluminescent maximum
spectrum. The hemispheric radiance due to the scattering of the
injected light by the photocyte structure and the extracting
structures examined in the present work was calculated from the
detailed data by integration over the measured solid angle. The
result is given by the red curve in Fig. \ref{fig11}. Then, the
same experiment was carried out after the surface of the segments
were coated with a wetting liquid chosen to match the refractive
index 1.56 of chitin. It was expected that the wetting liquid
reduced considerably the microscopic corrugation on the firefly
abdomen, so that the extracting advantage of these diffusers would
be significantly reduced. This is what has been actually observed
(see blue curve in Fig. \ref{fig11}). Note that only ratios of these values can be compared to the theoretical prediction, as the reference systems are optically slightly different. The extraction gain, due to
the corrugation wiped out by the liquid can be estimated to be
above 7\% at 560 nm, which means that some improvement of the
light extraction is actually observed. In this measurement we consider no change of the refractive index, only the addition (or deletion) of the specific tilted-scale structure. Therefore, the comparision has to be done with the simulations of substructure (3), i.e. to a light extraction gain of 9.5\% in comparision to a plane surface (see section \ref{computer}). The measurement approves the simulation and gives an experimental support. The difference of 2\% is acceptable considering that for the measurements, the actual light source is outisde of the abdomen and not inside the chitin, like in the simulations. The lantern of the dried specimen contains probably a lot of air spaces, which can influence the path of light.

\section{Conclusion}

The most effective mechanism found for the light extraction from
the lantern of fireflies is the diffuse transmission, produced by
the misfit of the scales and the resulting abrupt profile found on
the cuticle of the luminous ventral segments. Another, secondary,
improvement comes from the lowering of the refractive index at the
level of the photocytes, where the bioluminescence chemistry takes
place. The other structures found and analyzed do not contribute
to a better efficiency but are not detrimental.

These results, however, are only valid for the actual firefly
species we have examined. Many other insects produce their own
light and, for each of them, a careful analysis of their
morphology would be extremely useful. Fireflies and click-beetles,
for instance, do not all emit the same wavelength and it would be
interesting to know whether some correlation can be found between
the emitted wavelength and the optimized structures. With only one
example, it is obviously too early to dare any generalization.

The problem of light extraction in fireflies is strongly
correlated with many difficulties that must be faced when
developing artificial light sources from the solid state. As in
many other fields, a biomimetic approach might be helpful when
addressing these questions. The lowering refractive index of the
active emitting region and the introduction of a specific
diffusing roughness on the external surface of the device are
reasonable steps to be taken when considering the improvement of
solid-state light sources.

\section*{Appendix: Materials and Methods}

In this section, we briefly introduce experimental techniques and simulation highlights used in the present work. 

The specimens needed for the present study were all collected in
the Darien forest, at the Cana Field Station, in the back country
of the Republic of Panama. The fireflies were found flashing
mainly on the grassy air strip which gives access to the station
and on bordering trees. The \textbf{collection} started near 9 pm, on a
clear moonless night, early in May.

A \textbf{taxonomic determination} of the collected fireflies was attempted
by comparing our specimens with those kept in museums collections.
The section of Lampyridae from the Royal Belgian Institute of
Natural Sciences in Brussels and of the National Natural History
Museum of the Smithsonian Institution (Washington D.C.) were
examined and the subfamily -- Photurinae -- and the genus --
\textit{Photuris} (LeConte, 1851) -- could be determined. The
exact species was doubtful. Close matches, on external traits,
included \textit{trilineata}, a species that has been encountered
in Panama or close Colombian areas.

The specimens used in this study were examined with \textbf{Scanning Electron Microscopy} (SEM). For this investigation, small areas from the ventral side of the
abdomen were cut in liquid nitrogen and mounted, cross-section
face up, on a sample holder. Some of the samples were metallized
to ease charge elimination (20 nm of gold) and examined with two
scanning electron microscopes (SEM). A low-resolution Philips XL20
was chosen for preliminary explorations and a JEOL 7500F
high-resolution field-effect SEM was used for more detailed
examinations. Non-metallized samples were examined under a \bs
gentle beam'' mode where electrons decelerate before touching the
sample surface, avoiding charge build up.

SEM analysis is a two-dimensional surface analyzing technique: to be able to see the region of interest (ROI), this one has to be accessible to the electrons. That means, in the case of the firefly, we have to cut the lantern open to see the inside. \textbf{Nanotomography} in contrary is a three-dimensional analyzing technique, where the whole lantern can be scanned. Preparation of the samples included embedding in epoxy resin to stabilize spatially the sample. Other samples could be mounted immediately on the experimentation stage, without any preparation. The experiment has been carried out on the micro-imaging beamline \textit{ID22} at the European Synchrotron Radiation Facility (ESRF). Holotomography has been used on our samples, i.e. quantitative phase tomography with a coherent hard synchrotron radiation x-ray beam \cite{Cloetens-APL-1999}. Holotomography is used when the difference in density between components is weak, as in the firefly case. The sample is rotated 360$^\circ$ and two-dimensional projection images are taken at numerous angle positions. This procedure was repeated for three additional and well defined distances which leads to three-dimensional reconstructions.

Reflectivity and transmittivity spectra are calculated with \textbf{three-dimensional transfer-matrices} \cite{Forsythe-bk-1977}, a computational technique designed to describe the propagation of light in a photonic-crystal film from Maxwell's equations. This technique produces the intensity, at a fixed frequency, of each forward or backward diffracted beam from an inhomogeneous dielectric film with a two-dimensional lateral periodicity surfaces. Note that in this approach, the refractive index is not averaged laterally over periodic planes, so that both the lateral and normal inhomogeneities are fully accounted for. Technically, the transfer-matrix approach uses the same basis as the RCWA (rigorous coupled-wave analysis) technique used for calculating the intensity of diffracted waves on gratings \cite{Moharam-TM}. A one-dimensional, fully vector, version valid for a general planar multilayer, and a full three-dimensional version of this algorithm are available and used for the calculations needed in all simulations. The algorithm is highly stable for both one- and three-dimensional cases, thanks to the special way of constructiong the final scattering matrix, which relates all outgoing waves to given ingoing wave. In three-dimensional cases, the final scattering matrix results from the fine slicing of the structured film, the construction of the scattering matrix for each slice from the solution of Maxwell's equation and the analytic assembly of all scattering matrices. The calculation is essentially exact, except for the limited number of terms in the Fourier representation of the periodic lateral parts of the fields and dielectric functions and a staircase assumption, which considers that the refractive index is constant in each fine slice in the normal direction.

The radiance measurements have been carried out on a \textbf{scatterometer EZ Contrast XL80MS} built by the French company ELDIM. This instrument is essentially a system of Fourier lenses that redirect all light rays emitted along all emergence directions in such a way that they form an image of the source on a plane (called the \bs Fourier plane''). Each direction, characterized by a polar angle and an azimuthal angle is mapped to a point on a disk on the Fourier plane. The image is then transferred to a CCD captor. While being transferred, the light is filtered in order to pass only a small range of wavelengths (about 10 nm of specral width window). The captor measures all the intensities in the disk-shaped map which contains a planar projection of the intensity distribution in the emergence hemisphere. 31 colored filters, covering the visible region, allow to produce a complete relevant spectrum with a reasonable wavelength resolution. The result is the radiance (in units of W/m$^2$/nm/sr) finely analyzed angularly over a complete range (2$\pi$) of azimuthal angles and a nearly complete (from 0$^\circ$ to 80$^\circ$) range of polar angles, for 31 wavelengths covering the visible spectrum. This instrument is used here to give a direct measurement of the the light scattered by the photocytes of a firefly illuminated by light brought into the firefly lantern, through an optical fiber (diameter 200 $\mu$m), from an external white source. The source used for internal illumination is 300W EZ Reflex Source lamp which is able to deliver a maximal luminous emittance of 3000 lx. The radiance was integrated over all emergence angles to produce an absolute value of the emitted power per unit surface and unit  wavelength.

\section*{Acknowledgments}
We thank Dr. Jean-Fran\c{c}ois Colomer for his help in the
scanning-electron microscopes operation. J.P.V. thanks Dr. Alan R.
Gillogly, Curator of Entomology at the Orma J. Smith Natural
History Museum, Caldwell, Idaho, USA, and Dr. Donald Windsor for
help in collecting the specimens of fireflies used in the present
study. Both authors acknowledge very helpful discussions with
Prof. Helen Ghiradella (State University of New York, Albany,
USA), Dr. Laure Bonnaud (Museum National d'Histoire Naturelle,
Paris France), Prof. Laszl\`{o} Bir\`{o} (Nanostructures
Department MFA, Research Institute for Technical Physics and
Materials Science MTA, Hungarian Academy of Sciences, Budapest),
Dr. J\'{e}r\^{o}me Loicq (Li\`{e}ge Space Center, Belgium), Dr.
Pol Limbourg (Royal Belgian Institute of Natural Sciences,
Brussels, Belgium), Dr. Gary Hevel and Dr. Warren Steiner
(National Museum of Natural History, Smithsonian Institution,
Washington D.C., USA).The X-ray nanotomography experiment was carried out at the nano-imaging endstation
ID22NI of the European Synchrotron Radiation Facility
(ESRF) in Grenoble (France). A. B. was supported as PhD
student by the Belgian Fund for Industrial and Agricultural
Research (FRIA). The project was partly funded by the Action de
Recherche Concert\'{e}e (ARC) Grant No. 10/15-033 from the French
Community of Belgium. The authors also acknowledge using resources
from the Interuniversity Scientific Computing Facility located at
the University of Namur, Belgium, which is supported by the
F.R.S.-FNRS under convention No. 2.4617.07.

\end{document}